\documentclass[conference]{IEEEtran}
\IEEEoverridecommandlockouts

\usepackage{tikz}
\usepackage{cite}
\usepackage{amsmath}
\usepackage{newtxmath}

\usepackage{algorithmic}
\usepackage{graphicx}
\usepackage{makecell}
\usepackage{multirow}
\usepackage{threeparttable}
\usepackage{booktabs}
\usepackage{hyperref}

\usepackage{url}

\ifCLASSOPTIONcompsoc
\usepackage[caption=false, font=normalsize, labelfont=sf, textfont=sf]{subfig}
\else
\usepackage[caption=false, font=footnotesize]{subfig}
\fi

\begin{document}

\title{Partially Synchronous BFT Consensus Made Practical in Wireless Networks

\thanks{
Corresponding author: Minghui~Xu (\href{mailto:mhxu@sdu.edu.cn}{mhxu@sdu.edu.cn})

This study was supported by the National Natural Science Foundation of China (No. 62302266, 62232010, U23A20302), the Shandong Science Fund for Excellent Young Scholars (No.2023HWYQ-008), and the project ZR2022ZD02 supported by Shandong Provincial Natural Science Foundation.}

}

\author{
	\IEEEauthorblockN{Shuo~Liu$^{\dag}$, Minghui~Xu$^{\dag}$, Yuezhou~Zheng$^{\dag}$, Yifei~Zou$^{\dag}$, Wangjie~Qiu$^\ddagger$, Gang~Qu$^\S$, Xiuzhen~Cheng$^{\dag}$}
	\IEEEauthorblockA{$^\dag$ School of Computer Science and Technology, Shandong University}
        \IEEEauthorblockA{$^\ddagger$ School of Computer Science and Engineering, Beihang University}
        \IEEEauthorblockA{$^\S$ Department of Computer Science, University of Maryland}
}

\maketitle

\begin{abstract}
Consensus is becoming increasingly important in wireless networks. Partially synchronous BFT consensus, a significant branch of consensus, has made considerable progress in wired networks. However, its implementation in wireless networks, especially in dynamic ad hoc wireless networks, remains challenging. Existing wireless synchronous consensus protocols, despite being well-developed, are not readily adaptable to partially synchronous settings. Additionally, reliable communication, a cornerstone of BFT consensus, can lead to high message and time complexity in wireless networks. To address these challenges, we propose a wireless communication protocol called ReduceCatch (Reduce and Catch) that supports reliable 1-to-N, N-to-1, and N-to-N communications. We employ ReduceCatch to tailor three partially synchronous BFT consensus protocols (PBFT, Tendermint, and HotStuff) for seamless adaptation from wired to ad hoc wireless networks. To evaluate the performance of the ReduceCatch-enabled consensus protocols, we develop a three-layer wireless consensus testbed, based on which we implement 20 distinct consensus protocols and measure their latency and throughput. The experimental results demonstrate the superiority of the ReduceCatch-based consensus protocol in terms of latency and throughput.
\end{abstract}


\section{Introduction}
\label{sed:intro}
Wireless network applications such as dynamic task allocation \cite{TaskAllocation}, collective map building \cite{MapBuilding}, obstacle avoidance \cite{Obstacle}, and unmanned aerial vehicle/robot search and rescue \cite{rescue}, often require reaching a consensus before the initiation of subsequent tasks. Unfortunately, Byzantine attacks can cause consensus failures in wireless environments, resulting in significant losses. To address this issue, Byzantine fault-tolerant (BFT) consensus protocols, which have been designed to ensure agreement in the presence of Byzantine behaviors, get increasingly crucial in wireless networks.

%
%
BFT consensus protocols are classified based on their network synchrony assumptions: synchronous, partially synchronous, and asynchronous. Synchronous protocols prioritize speed and determinism but require highly reliable networks. Partially synchronous protocols offer greater resilience to network delays by relaxing timing constraints, making them suitable for environments with moderate unreliability. Asynchronous protocols can tolerate the most challenging network conditions but rely on complex cryptography and are more intricate to design.
In dynamic ad hoc wireless networks, the partially synchronous model is commonly employed, where message delays are bounded after a global stabilization time (GST). Despite extensive research on partially synchronous BFT protocols like PBFT \cite{PBFT}, Tendermint \cite{Tendermint}, and HotStuff \cite{HotStuff} for wired networks, adapting them for wireless environments, particularly dynamic ad hoc networks, presents significant challenges.
As a matter of fact, in a wired network, the establishment of reliable and stable channels between any two nodes is facilitated by wires and advanced protocols such as TCP and TLS. But in most ad hoc wireless networks, all nodes share channels, which can lead to inefficiencies in consensus protocols due to packet collision, channel congestion, and channel under-utilization. In particular, implementing partially synchronous BFT consensus protocols in dynamic ad hoc wireless networks faces the following three challenges.

\textbf{[Challenge I]} 
Existing wireless synchronous consensus protocols \cite{reli, A2, WirelessPaxos} have exhibited advancements in synchronous networks.
These protocols, nevertheless, usually require highly accurate time synchronization mechanisms or consistently synchronous operations, which limits their adaption to partially synchronous networks. 
Synchronous transmission (ST) based communication protocols depend on constructive interference \cite{glossy} and capture effect \cite{chaos}, which necessitate strong time synchronization. For instance, Glossy \cite{glossy} requires that the time difference between concurrent IEEE 802.15.4 transmitters should not exceed 0.5 $\mu s$ to ensure constructive interference with a high probability. Efficient real-time communication protocols such as \cite{song2008wirelesshart,he2003speed} require the network to constantly maintain a synchronous state, which contradicts the partially synchronous network model. 

\textbf{[Challenge II]} 
Partially synchronous BFT consensus requires reliable communications to tolerate packet losses \cite{PBFT, Tendermint, HotStuff}.
In wireless networks, reliable communications can be realized by combining CSMA (Carrier Sense Multiple Access) or TDMA (Time Division Multiple Access) \cite{survey_MAC} with ACK (Acknowledgement) or NACK (Negative Acknowledgement) \cite{ACK/NACK}.
However, the CSMA approach, where many nodes contend for a channel, often leads to severe network congestion \cite{shrestha2014distributed}. 
TDMA is constrained to uniformly allocate time slots among all nodes, due to the lack of information regarding the identities of the active nodes \cite{ahmed2016sleep}. Consequently, it is susceptible to allocating time slots to inactive nodes, thereby leading to channel idleness. In the worst case, TDMA might allocate $\mathcal{O}(N)$ time slots, where $N$ is the total number of nodes, even when there exist only $\mathcal{O}(1)$ active nodes. 

\textbf{[Challenge III]} The absence of a practical testbed specifically designed for validating partially synchronous BFT consensus in ad hoc wireless networks hinders evaluation. Existing network simulators like NS-3 and OMNeT++ lack built-in support for integrating consensus components.
Simulation mechanisms for consensus study adopted by \cite{802.11_PBFT, BLOWN, wChain} are unable to accurately replicate real-world environments, thereby failing to mimic real-time variations in latency and packet loss rates as well as their unpredictability. Besides, practical testing environments employed in \cite{TEE-1, TEE-2, Turquois} lack support for dynamic networks. Wireless sensor network (WSN) testbeds utilized in \cite{reli, A2, WirelessPaxos} suffer from the drawbacks of short-range transmission protocols and limited resources, making them inadequate for evaluating large-scale or complex collective behaviors such as post-disaster drone search and rescue operations \cite{rescue}, which often require coverage over expansive areas. Resource-constrained nodes within a WSN testbed struggle painfully to execute complex algorithms such as consensus protocols, due to limited computational capabilities.

To tackle the aforementioned challenges, this paper introduces ReduceCatch (Reduce and Catch), which incorporates TDMA, CSMA, and NACK to facilitate efficient communication patterns specifically designed for partially synchronous BFT consensus in ad hoc wireless networks. Additionally, we present a three-layer wireless consensus testbed that is utilized for evaluating the performance of partially synchronous BFT consensus in such networks. In summary, this paper makes three main contributions: 

\begin{enumerate}

\item \textbf{MAC Layer Protocol.} We propose ReduceCatch, a wireless communication protocol that has been specifically designed for practical uses in partially synchronous networks. What sets it apart from synchronous communication protocols is that ReduceCatch does not rely on synchronous transmissions and can be easily adapted to the dynamic characteristics of ad hoc wireless networks. Especially, the protocol has been optimized for upper-layer communication patterns and consensus protocols. 

\item \textbf{Communication Patterns.} We provide an efficient solution to implement reliable 1-to-N, N-to-1, and N-to-N communications as building blocks of consensus protocols. This approach is more effective than directly combining CSMA or TDMA with ACK or NACK. Specifically, we improve the efficiency of communication patterns in terms of both message\footnote{Message complexity denotes the total number of messages transmitted by all nodes throughout the execution of an algorithm.} and time complexity\footnote{Time complexity is the cumulative number of time slots consumed during the execution of an algorithm.}. This makes it an ideal choice for fast consensus protocols. A detailed comparison study on CSMA and TDMA combined with ACK and NACK as well as the flooding-based approach is presented in Table~\ref{tab:compare}.

\item \textbf{Consensus Protocols and Testbed.} We have developed and released a wireless consensus testbed that comprises three modular layers: physical layer, network layer, and consensus layer. This testbed is specifically designed to facilitate ad hoc wireless networks that are highly dynamic, such as swarms of smart cars, and it has extensive coverage using the LoRa technique. We have implemented a comprehensive suite of 20 diverse consensus algorithms using five distinct communication protocols. Our modular testing platform is intended to assist in improving protocol performance across multiple layers. Our code is open-sourced at \url{https://github.com/ReduceCatch/WirelessConsensus}.

\end{enumerate}


\section{Related Work}
\label{sec:related work}

\textbf{Byzantine Fault-Tolerant Wireless Consensus.} 
BFT wireless consensus protocols employ various techniques, including randomization \cite{Turquois}, Trusted Execution Environment (TEE) \cite{TEE-1, TEE-2}, spanner \cite{wChain}, carrier sensing, and random timer, to handle collisions \cite{802.11_PBFT}. For instance, by deeply integrating wireless communication characteristics with carrier sensing, the Proof-of-Channel wireless consensus protocol was proposed in \cite{BLOWN} for single-hop ad hoc wireless networks. A pioneering work \cite{bohm2024tinybft} deploys PBFT on resource-constrained ESP32.
Besides, BFT wired consensus can be transitioned from wired to wireless networks using traditional wireless communication protocols. For example, Goyal \textit{et al.} \cite{reli} explored the use of Glossy \cite{glossy} and Chaos \cite{chaos} to migrate PBFT to static and low-power wireless networks.
Additionally, general MAC protocols such as CSMA and TDMA \cite{BMAC, SMAC} can be utilized for high-dynamic applications without the reliance on synchronous transmissions. 
Specifically, utilizing ACK or NACK \cite{ACK/NACK}, these MAC protocols can be used to establish reliable communication patterns abstracted as CSMA-ACK, CSMA-NACK, TDMA-ACK, and TDMA-NACK.

\textbf{IoT Testbed.}
OMNeT++\footnote{https://github.com/omnetpp/omnetpp}, a tool for simulating wired and wireless network scenarios, does not support the required components of consensus.
Utilizing 802.11, Xu \textit{et al.} \cite{TEE-1, TEE-2} and Moniz \textit{et al.} \cite{Turquois} built static ad hoc wireless networks.
%
Flocklab2 \cite{FlockLab2} and FIT IoT-LAB \cite{FIT-IoT-LAB} are two publicly available wireless sensor network testbeds. 
FlockLab2 exhibits high synchronization accuracy, with an average synchronization error of 39$\mu s$, making it well-suited for the deployment of ST-based communication protocols \cite{glossy, chaos}. Nevertheless, all deployed nodes on FlockLab2 are static. FIT IoT-LAB offers 117 mobile robots, but it employs the 802.15.4 communication protocol, rendering it unsuitable for extensive coverage with a limited number of nodes.

 

\section{Models}
\label{sec:model}

\textbf{Network Model.} We consider an ad hoc wireless network that can be either single-hop or multi-hop. In the single-hop version, a set $V$ of $N$ mobile nodes can communicate with each other directly, while in the multi-hop version, nodes are partitioned into distinct clusters based on their geographic locations. Each cluster forms a single-hop network, with the $i^{th}$ cluster consisting of a set $V_i$ of $N_i$ mobile nodes. Each cluster has a randomly elected and changeable cluster leader, and communications between nodes at different clusters are facilitated through the cluster leaders.
Each node, whether in a single-hop or a multi-hop network, is equipped with a half-duplex transceiver, assigned a unique ID, and aware of the network/cluster size at the beginning of the protocol. 

We adopt a partially synchronous model \cite{partially_synchronous}, where the network alternates between asynchronous and synchronous states within a single-hop network (a cluster). In an asynchronous state, a global standardization time (GST) is incurred. After this period, the network transitions to a synchronous state. This assumption of partial synchrony is reasonable for an actual ad hoc wireless network, as continuously maintaining synchronous network operations is infeasible due to node mobility, which may make a node move out of the single-hop communication range, and the external environment, which often causes a high packet loss rate.  

\textbf{Adversary Model.} The nodes are categorized as either honest or Byzantine, with the former faithfully following a protocol, and the latter capable of launching any form of attacks. In a single-hop network, the number of Byzantine nodes is up to $f$, with $N=3f+1$; while in a multi-hop network, the number of Byzantine nodes is up to $f_i$, with $N_i=3f_i+1$ in the $i^{th}$ cluster. 
ReduceCatch aims to improve communication efficiency in partially synchronous BFT consensus by addressing aspects of both the lower MAC layer and higher-level communication patterns. Similar to existing BFT algorithms (e.g., PBFT~\cite{PBFT}, Tendermint~\cite{Tendermint}, HotStuff~\cite{HotStuff}), we focus on Byzantine behavior above the network layer, such as sending incorrect messages or collaborating with other Byzantine nodes. On the one hand, the BFT consensus process can effectively manage nodes deviating from the network-layer protocol, successfully countering adversaries if the number of Byzantine nodes is below the security threshold.
On the other hand, there is an orthogonal set of works \cite{pirayesh2022jamming, basu2020security} to counter various attacks on the network layer. 
Additionally, we rely on the security of cryptographic primitives used in this paper. Our work primarily focuses on performance enhancements and communication-level design, without modifying the core consensus procedures that ensure security, thus we do not include security proofs for brevity, as they can be found in previous papers.

\begin{table*}[!htb]
\centering
\begin{threeparttable}
\caption{Comparison of message complexity and time complexity in dynamic networks}
\label{tab:compare}
\tabcolsep=0.2cm
    \setlength{\tabcolsep}{2pt}{
    \begin{tabular}{l c l l l l l l }
    
        \toprule[1pt]
        
        \multicolumn{1}{l}{\multirow{2}{*}{}} 
        &\multicolumn{1}{c}{\multirow{2}{*}{Reliability}} 
        & \multicolumn{2}{c}{1-to-N}   
        & \multicolumn{2}{c}{N-to-1}     
        & \multicolumn{2}{c}{N-to-N}   \\
        
        \multicolumn{1}{l}{}      
        & \multicolumn{1}{c}{}
        & \multicolumn{1}{c}{\makecell{Message \\ Complexity}} 
        & \multicolumn{1}{c}{\makecell{Time \\ Complexity}}
        & \multicolumn{1}{c}{\makecell{Message \\ Complexity}} 
        & \multicolumn{1}{c}{\makecell{Time \\ Complexity}}
        & \multicolumn{1}{c}{\makecell{Message \\ Complexity}} 
        & \multicolumn{1}{c}{\makecell{Time \\ Complexity}} \\
        
        \midrule[0.5pt]
        Flooding-based    & $\times$     & $\mathcal{O}(1)$    & $\mathcal{O}(1)$     & $\mathcal{O}(N)$   & $\mathcal{O}(N)$  & $\mathcal{O}(N)$    & $\mathcal{O}(N)$  \\
        CSMA-ACK         & \checkmark  & $\mathcal{O}(N\log N)$    & $\mathcal{O}(\beta N\log (\beta N))$    & $\mathcal{O}(N\log N)$   & $\mathcal{O}(\beta N\log (\beta N))$  & $\mathcal{O}(N^2\log N)$  & $\mathcal{O}((\beta N)^2\log (\beta N))$ \\
        CSMA-NACK        & \checkmark  & $\mathcal{O}(N)$    & $\mathcal{O}(\beta N)$    & $\mathcal{O}(N\log N)$   & $\mathcal{O}(\beta N\log (\beta N))$  & $\mathcal{O}(N^2)$    & $\mathcal{O}((\beta N)^2) $ \\
        TDMA-ACK         & \checkmark  & $\mathcal{O}(N\log N)$    & $\mathcal{O}(N\log N)$    & $\mathcal{O}(N\log N)$   & $\mathcal{O}(N\log N + N)$  & $\mathcal{O}(N^2\log N)$  & $\mathcal{O}(N^2\log N)$ \\
        TDMA-NACK        & \checkmark  & $\mathcal{O}(N)$    & $\mathcal{O}(N\log N)$    & $\mathcal{O}(N\log N)$   & $\mathcal{O}(N\log N+N)$  & $\mathcal{O}(N^2)$    & $\mathcal{O}(N^2) $ \\
        \textbf{ReduceCatch}         & \checkmark  & $\boldsymbol{\mathcal{O}(\log N)}$    & $\boldsymbol{\mathcal{O}(\log N)}$    & $\mathcal{O}(N\log N)$  & $\boldsymbol{\mathcal{O}(N\log N + \beta n)}$  & $\boldsymbol{\mathcal{O}(N\log N)}$   & $\boldsymbol{\mathcal{O}(N\log N)}$\\
        \bottomrule[1pt]
    \end{tabular}}
    \begin{tablenotes}
        \footnotesize
        \item[$n$] Remaining active nodes after the reduce phase. $\mathcal{O}(n)$ can be $\mathcal{O}(1)$, when $\mathsf{NTX}=\log N$ in all three communication patterns.
        \item[$\beta$] Expansion coefficient used in CSMA, caused by packet collisions. In the presence of $\mathcal{O}(N)$ active nodes, it can be $\mathcal{O}(1)$ or $\mathcal{O}(N)$, depending on different CSMA details.
      \end{tablenotes}
    
\end{threeparttable}
\end{table*}

\section{Design of ReduceCatch}
\label{sec:design}

\subsection{Limitations of Traditional Communication Protocols for Consensus}
\begin{figure}[!htb]
    \centering
    \includegraphics[width=0.47\textwidth]{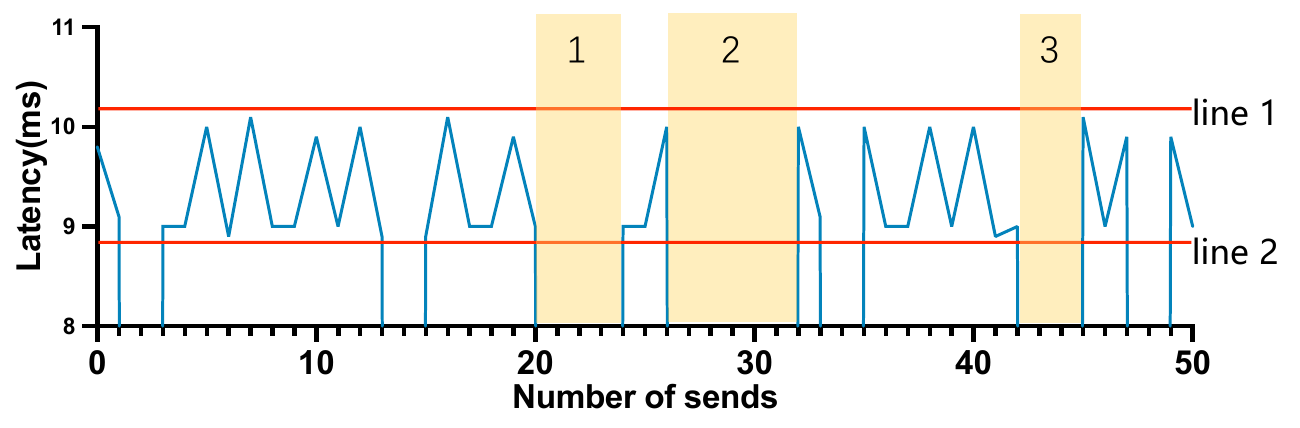}
    \caption{An example of packet loss}
    \label{fig:casestudy}
    
\end{figure}

Traditional communication protocols can be categorized into two categories, namely flooding-based, which inherently lacks reliability guarantees, and abstract communication protocol-based, which ensures message reliability.  
Flooding-based protocols may result in complete message losses for a sender as they allocate multiple consecutive slots to a node. This is because the packet loss rate frequently becomes high from time to time in wireless networks, especially for long-range communications, as the corresponding protocols (e.g., LoRa) are highly susceptible to environmental interference, particularly in the absence of signal reinforcement from base stations \cite{LoRa-1, LoRa-2}.
For illustration we present an instance of communication between two nodes employing LoRa in Fig.~\ref{fig:casestudy}, where a transmitter continuously sends messages to a receiver. The two nodes are about 300 meters apart. The $x$-axis denotes the number of message transmissions, while the $y$-axis indicates the delay between the receptions of two consecutive messages by the receiver. A delay that surpasses line 1 or is below line 2 indicates a message loss. Successive packet losses occur during periods 1, 2, and 3. Assigning these consecutive time slots to one node leads to a complete message loss for that node.

The four traditional abstract communication protocols, i.e., CSMA-ACK, CSMA-NACK, TDMA-ACK, and TDMA-NACK,  face the following major issues. 
Firstly, using CSMA alone leads to significant network congestion, whereas TDMA alone can easily allocate time slots to inactive nodes. When there are plenty of active nodes (especially in consensus protocols where all followers need to broadcast their votes), CSMA can cause severe network congestion due to the hidden terminal problem. Denote the network congestion in this context by the parameter $\beta$. One can see that
$\mathcal{O}(\beta)$ can be $\mathcal{O}(1)$ or $\mathcal{O}(N)$, depending on the CSMA protocol details and the number of active nodes.
Specifically, in a N-to-N communication, CSMA-ACK results in $\mathcal{O}(N^2\log N)$ message complexity and $\mathcal{O}((\beta N)^2\log (\beta N))$ time complexity, while CSMA-NACK results in $\mathcal{O}(N^2)$ message complexity and $\mathcal{O}((\beta N)^2)$ time complexity. The other two communication patterns also have similar message complexity and time complexity, as reported in Table~\ref{tab:compare}.

Secondly, in the absence of knowledge regarding the active status of an individual node, TDMA is constrained to uniformly distribute time slots among all nodes.  If the number of active nodes decreases as the protocol runs, TDMA is prone to allocating time slots to more inactive nodes, resulting in stronger channel idleness. Even there are $\mathcal{O}(1)$ active nodes, TDMA still needs $\mathcal{O}(N)$ time slots.
Therefore, TDMA-ACK or TDMA-NACK can all result in $\mathcal{O}(N\log N + N)=\mathcal{O}(N\log N)$ time complexity in a N-to-1 communication. The other two communication patterns utilizing TDMA-ACK or TDMA-NACK have the same issue as the N-to-1 communication. Note that for simplicity and convenient comparison with ReduceCatch, we put $\mathcal{O}(N\log N + N)$ instead of $\mathcal{O}(N\log N)$ in the N-to-1 communication in Table~\ref{tab:compare}.

Moreover, the utilization of ACK or NACK directly on top of CSMA or TDMA can result in high message complexity. 
The ACK mechanism necessitates feedback from all target receivers, leading to high message complexity. Particularly, in a N-to-N communication, using ACK with CSMA or TDMA  results in $\mathcal{O}(N^2\log N)$ message complexity. 
Instead, NACK messages 
should only be transmitted by receivers who miss the needed message. When network conditions deteriorate to the point where wireless broadcasting resembles wired unicast, the NACK mechanism leads to high message complexity. In a N-to-N communication, this complexity reaches $\mathcal{O}(N^2)$, whether using CSMA or TDMA.
The other communication patterns have the same problem as that in the N-to-N communication as shown in Table~\ref{tab:compare}.

\subsection{ReduceCatch}
\label{sec:design:basic_idea}
We propose a wireless communication protocol called ReduceCatch, which incorporates TDMA, CSMA, and NACK to establish reliable communication patterns. The main concept behind ReduceCatch is a two-phased mechanism that utilizes the first phase (reduce phase) to reduce the number of active nodes and the second phase (catch phase) to guarantee the message reliability of the remaining nodes.

\textbf{Reduce Phase.} At the beginning of a consensus protocol, the large number of active nodes can pose a challenge to efficient communications. The reduce phase utilizes TDMA to allocate time slots to each sender. The number of time slots allocated to each sender is represented by a system parameter $\mathsf{NTX}$. 
Unlike flooding-based protocols, consecutive time slots are not allocated to the same node in the reduce phase. Instead, this phase assigns a time slot to each node, and cycles $\mathsf{NTX}$ rounds. 

The determination of the appropriate value for $\mathsf{NTX}$ can
decrease the message and time complexities of communication patterns, by reducing the number of active nodes in a network.
In principle, the value of $\mathsf{NTX}$ should depend on the packet loss rate $\alpha$ and the network scale $N$. We elucidate the theoretical process of establishing the value of $\mathsf{NTX}$ within a reliable N-to-N communication. For a given node $i$ to transition from active to inactive, two conditions must be held: 1) receiving data from the other $N-1$ nodes, and 2) the other $N-1$ nodes receiving data from node $i$. We assume each packet loss is independent. Therefore, the probability of holding condition 1 or 2 is $(1-\alpha ^ {\mathsf{NTX}})^{(N-1)}$. Consequently, the probability of the node $i$ transitioning from active to inactive is $(1-\alpha ^ {\mathsf{NTX}})^{(2N-2)}$. Therefore, after the reduce phase, the number of active nodes is $n=N-N(1-\alpha ^ {\mathsf{NTX}})^{(2N-2)}$. If $\mathcal{O}(n)$ is to be reduced to $\mathcal{O}(1)$, $\mathsf{NTX}$ needs to be set to $\mathcal{O}(\log N)$. Same conclusions can also be made for 1-to-N and N-to-1 communications. 


\begin{figure*} 
\centering
  \subfloat[1-to-N ($\mathsf{NTX}=3,\Delta=5$)]{
   \includegraphics[width=0.31\linewidth]{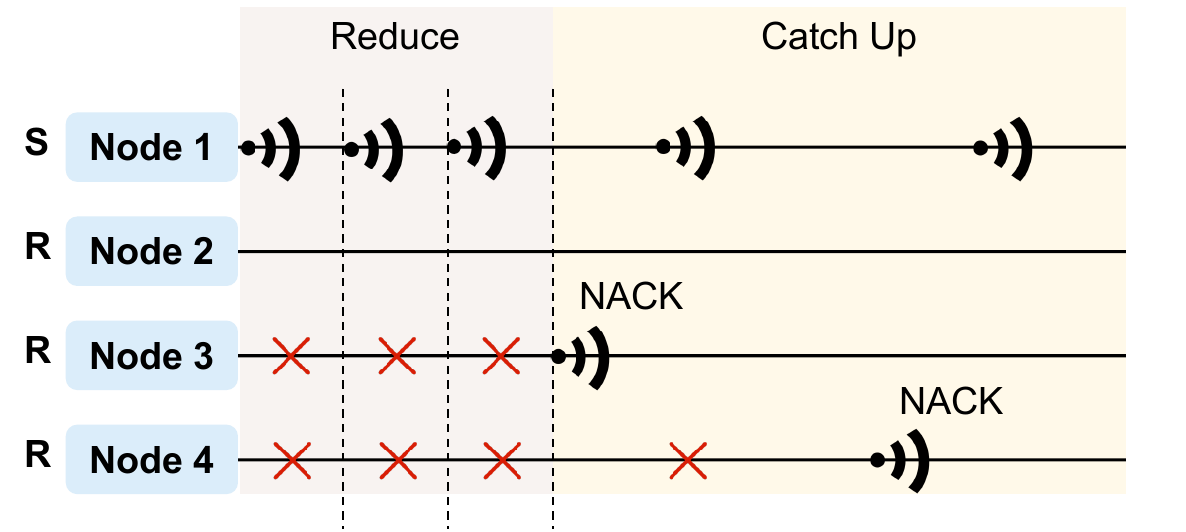}
\label{fig:1-to-N}}\hfill
  \subfloat[N-to-1 ($\mathsf{NTX}=2,\Delta=5$)]{
    \includegraphics[width=0.31\linewidth]{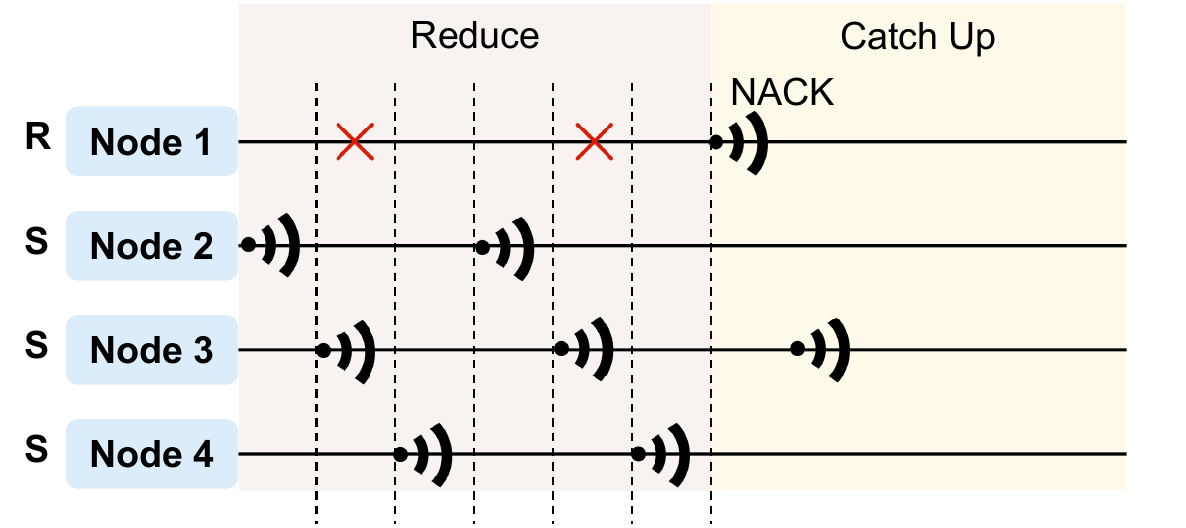}
\label{fig:N-to-1}}\hfill
  \subfloat[N-to-N ($\mathsf{NTX}=2,\Delta=6$)]{
    \includegraphics[width=0.31\linewidth]{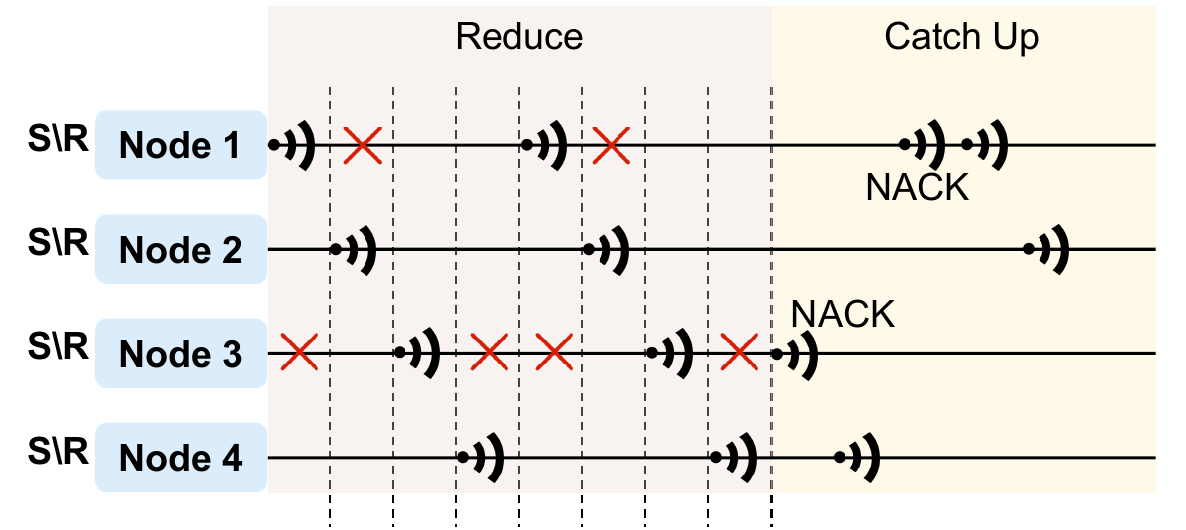}
\label{fig:N-to-N}}
  \caption{Communication patterns based on ReduceCatch. S denotes the sender and R denotes a receiver.}
  \label{fig:Different Communication patterns} 
\end{figure*}

\textbf{Catch Phase.} 
Although the reduce phase can decrease the number of active nodes to a constant magnitude, we still need the catch phase to ensure message reliability, mainly for two reasons. Firstly, key messages, such as proposals from a leader, need to be ensured to reach all honest nodes.
Secondly, as time goes, more and more nodes may miss messages, eventually leading to the failure of consensus.
Thus, we use CSMA and NACK in the catch phase to ensure message reliability for the remaining nodes from the reduce phase. These remaining nodes contend for the channel to transmit NACK messages to the target node. Upon receiving a NACK message, the target node resends its own message. 
To prevent Byzantine nodes from sending NACK messages maliciously, a timer $\Delta$ is utilized to restrict the acceptance of NACK messages outside a time-span. Besides, nodes can send NACK messages repeatedly to improve the probability of successful reception at the target node.

\subsection{Communication Patterns}
\label{sec:design:communication patterns}
We detail the use of ReduceCatch in establishing reliable communication patterns, i.e., 1-to-N, N-to-1, and N-to-N, which are essential in constructing consensus protocols such as PBFT \cite{PBFT}, Tendermint \cite{Tendermint}, and HotStuff \cite{HotStuff}. The 1-to-1 communication pattern is not considered, considering the wireless broadcasting nature in consensus scenarios.


\subsubsection{Communication Patterns based on ReduceCatch}
In a 1-to-N communication, a sender repeatedly broadcasts its data packet $\mathsf{NTX}$ times during the reduce phase. 
Although necessitating the allocation of multiple consecutive time slots to one node, increasing the value of $\mathsf{NTX}$ can alleviate the challenges associated with a high packet loss rate.
During the catch phase, nodes that miss messages transmit NACK, prompting the sender to resend its data packet. To prevent malicious broadcasting of NACK messages by Byzantine nodes, the catch phase has a time span of $\Delta$ time slots. Fig.~\ref{fig:1-to-N} illustrates an example, in which node 1 transmits data to nodes 2, 3, and 4, with $\mathsf{NTX}$ set to 3. Node 1 sends its data packet three times in the reduce phase, while the catch phase lasts $\Delta=5$ time slots, during which node 1 retransmits its data packet after receiving a NACK message.

In a N-to-1 communication, $N$ senders sequentially transmit their data packets during the reduce phase, which repeats for $\mathsf{NTX}$ cycles. The sole receiver transmits NACK messages during the catch phase to indicate which nodes' messages have been missed. Within a time span $\Delta$, senders who receive NACK messages retransmit their messages if the NACK messages indicate that the receiver has not received their messages. For example, in Fig.~\ref{fig:N-to-1}, nodes 2, 3, and 4 cyclically send their messages once during the reduce phase, looping $\mathsf{NTX}$ times. In the catch phase, node 1 transmits a NACK message to indicate which nodes' messages it did not receive. In the provided diagram, node 3's message is missed at node 1, so node 3 retransmits its message. The catch phase lasts $\Delta=5$ time slots to prevent malicious attacks from Byzantine nodes.

In a N-to-N communication, $N$ nodes sequentially transmit their data packets during the reduce phase, iterating $\mathsf{NTX}$ times. In the subsequent catch phase, nodes that have missed messages during the reduce phase transmit NACK messages, indicating which nodes' messages they have not received. This process continues for $\Delta$ time slots. Failing to complete communications within the $\Delta$ time limit is considered a failure. In complex communication patterns such as N-to-N, increasing $\Delta$ can improve the probability of completing the communication pattern. For example, in Fig.~\ref{fig:N-to-N}, four nodes need to collect messages from all other nodes. During the reduce phase, the four nodes sequentially transmit their messages, iterating $\mathsf{NTX}=2$ times. Node 3 does not receive messages from nodes 1 and 4, while node 1 does not receive a message from node 2. Upon receiving the NACK messages, nodes 1, 2, and 4 retransmit their packets. All nodes must await $\Delta=6$ time slots before proceeding to the next phase. 

\subsubsection{Advantages of Applying ReduceCatch}
ReduceCatch exhibits low message and time complexities in the construction of diverse communication patterns due to two reasons. Firstly, the reduce phase serves to decrease the number of active nodes. One can see from Section~\ref{sec:design:basic_idea} that when $\mathsf{NTX}$ is set to $\mathcal{O}(\log N)$, the reduce phase decreases the number of active nodes from $\mathcal{O}(N)$ to $\mathcal{O}(1)$. Therefore, utilizing ReduceCatch leads to  $\mathcal{O}(\log N)$ message complexity and $\mathcal{O}(\log N)$ time complexity for the 1-to-N communication, and $\mathcal{O}(N\log N)$ message complexity and $\mathcal{O}(N\log N)$ time complexity for the N-to-N communication.
Secondly, when the number of active nodes decreases, the catch phase makes use of CSMA instead of TDMA. The underlying reason for this is that TDMA necessitates $\mathcal{O}(N)$ time slots for even $\mathcal{O}(1)$ active nodes, due to the lack of prior information about the identities of the active nodes. Therefore, in a N-to-1 communication, ReduceCatch results in $\mathcal{O}(N\log N +\beta n)$ time complexity, where $\mathcal{O}(\beta n)$ can be $\mathcal{O}(1)$\footnote{When the value of $\mathsf{NTX}$ is set to $\mathcal{O}(\log N)$, $\mathcal{O}(n)$ becomes $\mathcal{O}(1)$. Besides, $\mathcal{O}(\beta)$ can be $\mathcal{O}(1)$, when using specific CSMA protocol.}, while
TDMA-ACK or TDMA-NACK results in $\mathcal{O}(N\log N + N)$ time complexity. 

\section{Wireless Consensus Protocols}
\label{sec:wireless consensus}
This section focuses on the usage of ReduceCatch to adjust three partially synchronous BFT consensus protocols -- PBFT, Tendermint, and HotStuff, to make them practically adaptable to both single-hop and multi-hop ad hoc wireless networks. 

\subsection{Single-Hop}
The communication patterns discussed in Section~\ref{sec:design:communication patterns} cannot be directly applied to wireless consensus; further integration is necessary. Firstly, the terminal conditions for each phase vary across different consensus protocols, requiring appropriate adjustments to accommodate these differences. Secondly, when employing multiple communication patterns consecutively, further optimization is needed. 

In the first phase, all three wireless consensus protocols involve the leader disseminating its proposal to all followers, necessitating a 1-to-N communication. The subsequent phases differ for each protocol. In Wireless PBFT, during the next two phases, each node broadcasts its vote and collects $2f+1$ matching votes for the same proposal at each phase. Thus, a N-to-N communication is necessary. In Wireless HotStuff, during the next three phases, the leader first collects a set of $2f+1$ matching votes from the previous phase and then broadcasts this set to all followers, necessitating N-to-1 and 1-to-N communications. In Wireless Tendermint, during the next two phases, it can utilize N-to-N communication, similar to Wireless PBFT, where each node broadcasts its vote, and each node independently collects votes. Alternatively, it can utilize N-to-1 and 1-to-N communications, similar to Wireless HotStuff, where the leader is responsible for collecting votes and broadcasting them.

The view change phase differs among the three wireless consensus protocols. Wireless PBFT only changes leaders when the current leader is malicious, requiring all nodes to broadcast their state, necessitating N-to-N communication. Wireless Tendermint and Wireless HotStuff periodically change leaders during normal operation. In Wireless Tendermint, the new leader must wait until all followers' \textit{view change} messages have been collected before proceeding to the next view. If the new leader cannot collect all followers' \textit{view change} messages, it must wait for at most GST time. In Wireless HotStuff, the new leader only needs to collect $2f+1$ matching \textit{view change} messages for the same view. For safety considerations, Wireless HotStuff adds one more phase in normal operation compared to the other two consensus algorithms.

\textbf{Optimization.} 
When multiple communication patterns are utilized consecutively, they can be further integrated by merging their catch phases.
For instance, we can integrate 1-to-N and N-to-1 communications using one reduce phase and one catch phase. During the reduce phase, the first $\mathsf{NTX}$ time slots are allocated to the leader for broadcasting its message. Subsequently, all followers sequentially broadcast their votes, iterating $\mathsf{NTX}$ times. If a follower does not receive the leader's message, it should keep silent and send NACK messages in the following catch phase. During the catch phase, if the leader fails to collect $2f+1$ identical votes or followers do not receive the leader's message, they can send NACK messages to request a resend.

\subsection{Multi-Hop}
Applying ReduceCatch in multi-hop networks is challenging, as the multi-hop topology affects the efficiency of ReduceCatch. A common approach in such a network is to employ ReduceCatch as if it were a single-hop network, but with each node relaying newly received packets. However, such an approach poses two significant challenges. Firstly, in the reduce phase, multiple nodes may relay different packets, leading to collisions that can hinder the reduction of active nodes. Secondly, a large number of active nodes and the complex topology of multi-hop networks make it difficult to determine the value of $\Delta$ in the catch phase. To overcome these issues, we partition the network into clusters with each representing a single-hop network, and then apply ReduceCatch within each cluster. 

\begin{figure}[!htb]
    \centering
    \includegraphics[width=0.8\linewidth]{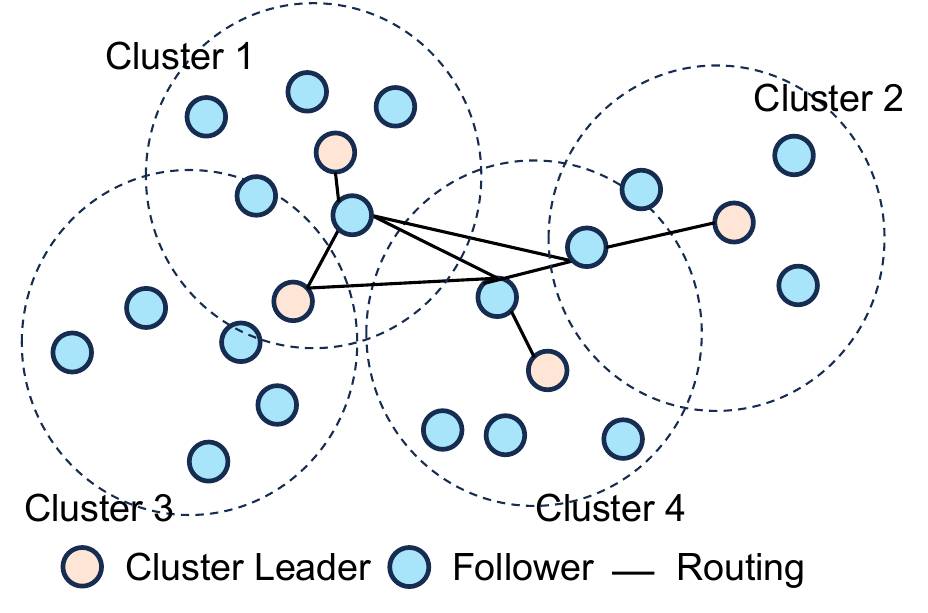}
    \caption{Multi-Hop topology}
    \label{fig:Multi-Hop Topology}
\end{figure}

Fig.~\ref{fig:Multi-Hop Topology} demonstrates a multi-hop topology that enables an efficient application of ReduceCatch. 
This solution approach adopts a parallelized design, similar to sharding in blockchain \cite{shard-1, shard-2}, in which many local consensus processes parallelly agree on different proposals, and rely on a global consensus to finally determine the order of these proposals. To proceed, the network is divided into multiple clusters, and each cluster is composed of nodes that can communicate with each other over one hop distance. Each cluster randomly elects a cluster leader\footnote{For simplicity, we refer to the consensus leader as the leader. The cluster leader can be different from the consensus leader to reduce the burden on a single node. } using any existing leader election algorithm \cite{leader-1, leader-2}, and this leader can be changed.
Local consensus is firstly achieved within each cluster, and distinct channels are employed within each cluster to prevent interference. Following the attainment of local consensus, the leader proceeds to participate in global consensus. Note that leaders involved in global consensus use a channel that is different from those adopted within each cluster to prevent interference. Communications between leaders are established via commonly employed routing protocols \cite{Routing-1,Routing-2} in wireless networks. Due to space constraints, we do not discuss the specific implementation of such protocols. Local consensus can ensure safety and liveness, as long as the number of Byzantine nodes does not exceed one-third of the total number of nodes in the cluster. Global consensus ultimately ensures that the majority of the leaders are honest, thereby guaranteeing safety and liveness, as local consensus ultimately selects an honest leader. Even if global consensus is controlled by Byzantine nodes, followers in local consensus can detect the incorrect proposal sequence and subsequently change the leaders.

\section{Wireless Consensus Testbed}
\label{sec:testbed}

\begin{figure} 
\centering
  \subfloat[]{
   \includegraphics[width=0.47\linewidth]{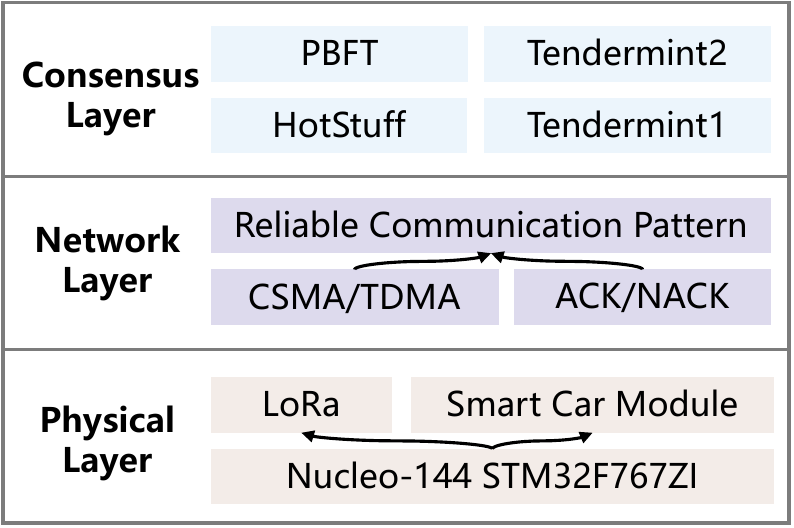}
\label{fig:wireless_consensus_testbed}}\hfill
  \subfloat[]{
    \includegraphics[width=0.47\linewidth]{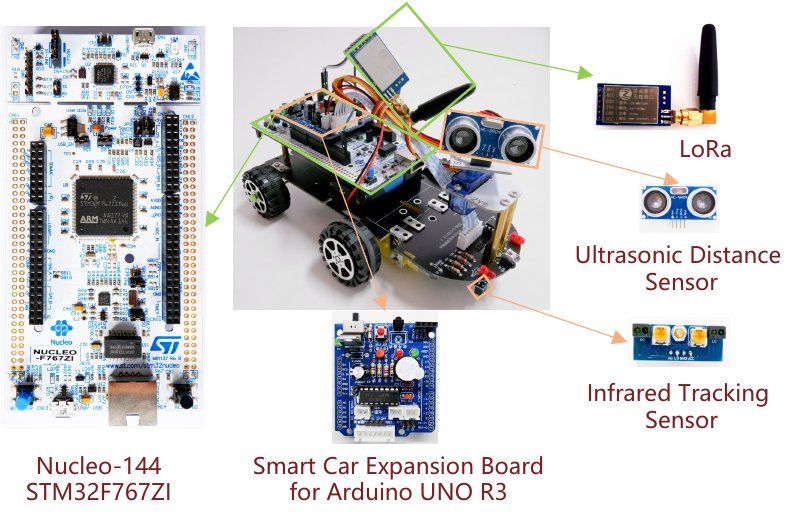}
\label{fig:smartcar}}
  \caption{Wireless consensus testbed. (a) Wireless consensus testbed architecture. The black arrows indicate the support relations. (b) Smart car parts breakdown.}
  \label{fig:Physical Layer} 
\end{figure}

To evaluate partially synchronous BFT consensus in practical ad hoc wireless networks, we develop a wireless consensus testbed, encompassing a physical layer, a network layer, and a consensus layer. Our testbed is designed in a modular fashion, which enables independence among the three layers and enhances their convenience for a seamless combination. We have made the codebase of the testbed publicly available, allowing other developers to rapidly construct their own systems on top of it.
%
Fig.~\ref{fig:wireless_consensus_testbed} presents the architecture of our proposed wireless consensus testbed.


\subsection{Physical Layer}

As demonstrated in Fig.~\ref{fig:wireless_consensus_testbed}, the physical layer of the testbed comprises three fundamental modules: a microcontroller unit (MCU), a communication module, and a smart car module. The corresponding components within a smart car is illustrated in Fig.~\ref{fig:smartcar}.



\textbf{MCU \& Communication Module.}
In order to support complex collective behaviors, MCUs should have enough computing power. For example, in our testbed, the physical layer's core component is the STM32 Nucleo-144 development board, which integrates the STM32F767ZI MCU\footnote{This MCU utilizes the Arm Cortex-M7 architecture, operating at a maximum frequency of 216 MHz, with 2 MB Flash and 512 KB SRAM storage capacity.} that delivers adequate performance to support intricate systems such as consensus. 
%
The communication module utilizes LoRa, which is a wireless communication technology that satisfies the requirements of low-power consumption and long-range transmission. Long-range transmission allows for the construction of single-hop systems in a wider range of application scenarios compared to short-range communication technologies. Single-hop systems are easier to deploy than multi-hop ones, making them a more attractive option. 
%
The LoRa employed in this testbed is based on the SX1268 chip, operating within a frequency range of 410MHz to 493MHz (in Asia), with a communication distance of up to 6km, and a maximum air rate of 62.5kbps. 

\textbf{Smart Car Module \& Smart Car Swarm.}
To support dynamic ad hoc wireless networking, the Smart Car module comprises mainly a Smart Car Expansion Board for Arduino UNO R3, an Ultrasonic Distance Sensor, and an Infrared Tracking Sensor. The expansion board facilitates the integration of diverse smart car components, including motors, Bluetooth, power supply, and multiple sensors, into the STM32 Nucleo-144 development board. The ultrasonic distance sensor employs ultrasonic waves to detect obstacles, thereby averting collisions. Similarly, the infrared tracking sensor deploys infrared rays to detect ground track lines for tracking purposes. 
The smart car swarm can be leveraged to simulate complex collective behaviors \cite{swarm_engineer}. 
Our testbed provides various antennas for LoRa, allowing for a maximum communication range that extends from 1 meter to 6 kilometers.
Additionally, all cars remain in perpetual motion, supporting a dynamic network.


\subsection{Network Layer and Consensus Layer}
The network layer consists of a MAC module, a reliability mechanism module, and a reliable communication pattern module.
The combination of the reliability mechanism module and the MAC module permits different implementation methods of the same reliable communication pattern. 
In addition to ReduceCatch, the provided codebase contains four direct combinations: CSMA-ACK, CSMA-NACK, TDMA-ACK, and TDMA-NACK. 
%
Additionally, we employ FreeRTOS\footnote{https://www.freertos.org/index.html}, a lightweight real-time operating system that occupies only 9 KB of memory in our testbed. FreeRTOS offers a variety of features and APIs, including task management, event management, and memory management, making it easier to design and develop consensus protocols. 

The consensus layer includes three consensus protocols: PBFT, Tendermint (Tendermint1 and Tendermint2), and HotStuff. 
These consensus protocols can be implemented using different forms of reliable communication patterns at the network layer. For example, in PBFT, two communication patterns are utilized: 1-to-N (PRE-PREPARE) and N-to-N (PREPARE, COMMIT). Each communication pattern can choose its own implementation, which is selected from ReduceCatch, CSMA-ACK, CSMA-NACK, TDMA-ACK, and TDMA-NACK. Developers are also allowed to design their own protocols. 
%
Consensus uses Micro-ECC\footnote{https://github.com/kmackay/micro-ecc} for message signing and verification. This lightweight library offers essential public key cryptography for resource-constrained embedded systems by avoiding dynamic memory allocation, a potential source of memory overflows.


\begin{figure*} 
\centering
   \includegraphics[width=0.5\textwidth]{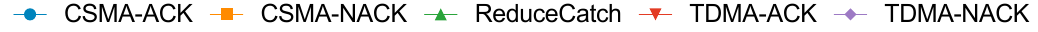}
    \\
  \subfloat[PBFT]{
   \includegraphics[width=0.23\linewidth]{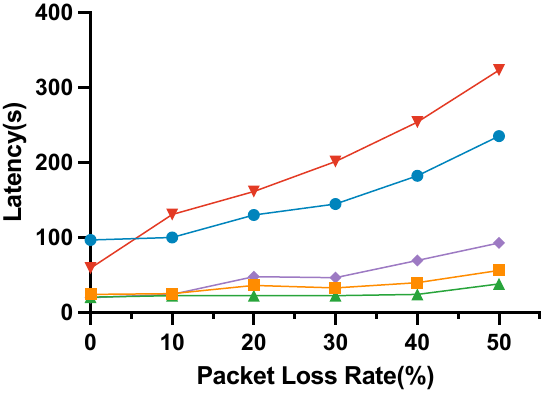}
\label{fig:Dynamic-PBFT}}\hfill
\subfloat[Tendermint1]{
   \includegraphics[width=0.23\linewidth]{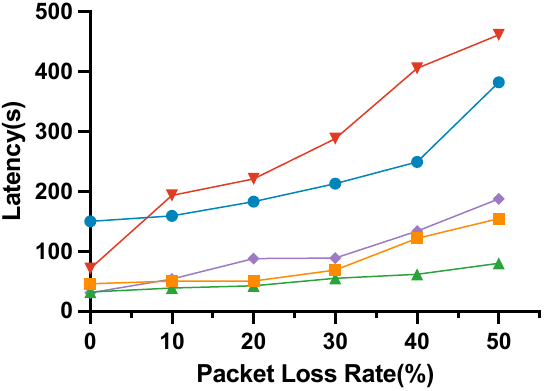}
\label{fig:Dynamic-Tendermint1}}\hfill
\subfloat[Tendermint2]{
   \includegraphics[width=0.23\linewidth]{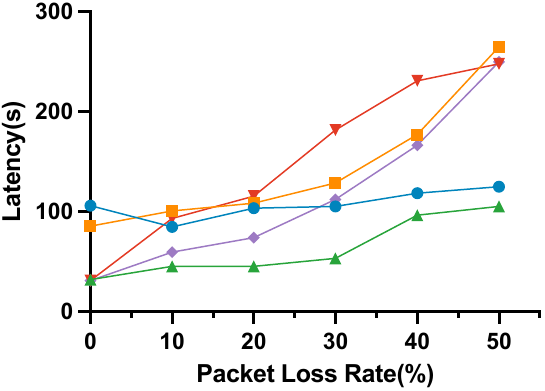}
\label{fig:Dynamic-Tendermint2}}\hfill
  \subfloat[HotStuff]{
    \includegraphics[width=0.23\linewidth]{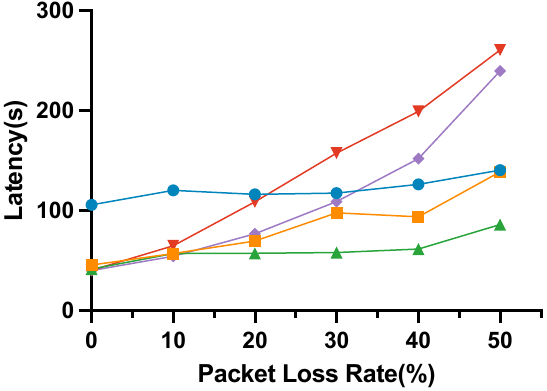}
\label{fig:Dynamic-HotStuff}}
  \caption{The latency of four consensus algorithms vs. network dynamism}
  \label{fig:Dynamic-Latency} 
\end{figure*}


\begin{figure*} 
\centering
   \includegraphics[width=0.5\textwidth]{img/data/Legend}
    \\
  \subfloat[PBFT]{
   \includegraphics[width=0.23\linewidth]{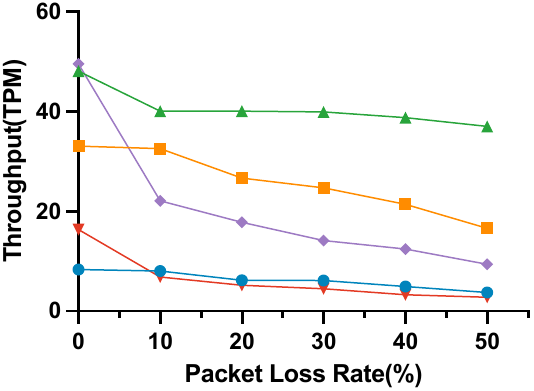}
\label{fig:Throughput-Dynamic-PBFT}}\hfill
\subfloat[Tendermint1]{
   \includegraphics[width=0.23\linewidth]{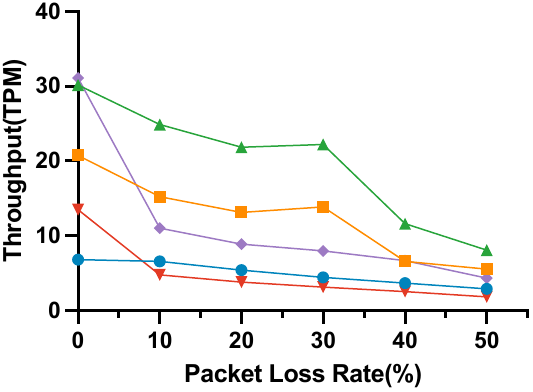}
\label{fig:Throughput-Dynamic-Tendermint1}}\hfill
\subfloat[Tendermint2]{
   \includegraphics[width=0.23\linewidth]{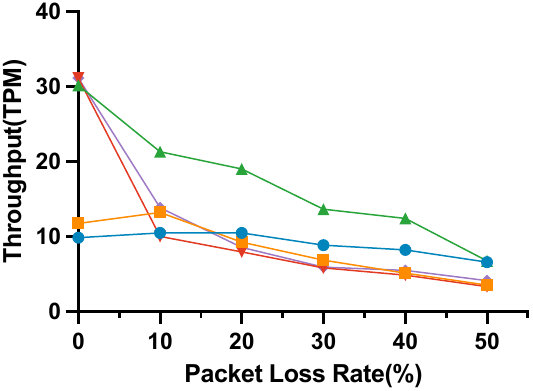}
\label{fig:Throughput-Dynamic-Tendermint2}}\hfill
  \subfloat[HotStuff]{
    \includegraphics[width=0.23\linewidth]{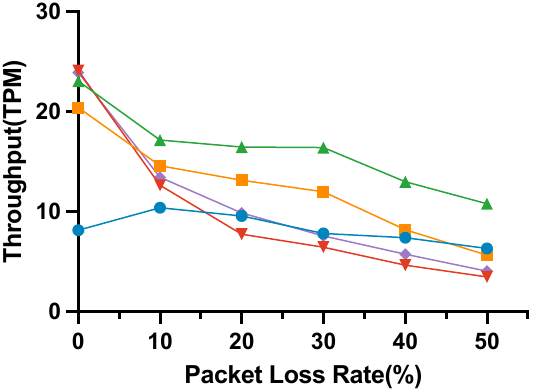}
\label{fig:Throughput-Dynamic-HotStuff}}
  \caption{The throughput of four consensus algorithms vs. network dynamism}
  \label{fig:Throughput-Dynamic-Latency} 
\end{figure*}

\section{Evaluation}
\label{sec:evaluation}

The wireless consensus testbed, introduced in Section~\ref{sec:testbed}, is utilized to conduct experiments on five distinct protocols, namely CSMA-ACK, CSMA-NACK, TDMA-ACK, TDMA-NACK, and our proposed ReduceCatch protocol. Flooding-based protocols are not considered because they have no reliability guarantee. For illustrative clarity, we focus on the core CSMA and TDMA mechanisms, acknowledging that enhancements exist. Three consensus algorithms, namely PBFT, Tendermint (Tendermint1 and Tendermint2), and HotStuff are deployed, resulting in a combination of 20 diverse consensus protocols. The total code size of the tested system in this section is 40,316 lines of C/C++ code. Five measurements are conducted for each sample point and the mean is calculated. Our testing effectively addresses the following questions: 
1) Under different degrees of network dynamics (Section~\ref{sec:evaluation:Dynamic}) or multi-hop networking conditions (Section~\ref{sec:evaluation:Multi-Hop}), what level of improvement is offered in terms of latency and throughput by the ReduceCatch compared to other communication protocols for the same consensus algorithm? 2) What is the performance of the three consensus algorithms in terms of latency and throughput? 

\subsection{Latency and Throughput}
\label{sec:evaluation:Dynamic}
We aim to evaluate the benefits of the ReduceCatch protocol in dynamic networks by comparing it to other communication protocols. 
For this purpose, we set the packet loss rate $\alpha = 0:10\%:50\%$ and the total number of nodes $(N)$ to 10. We also establish the values for $\mathsf{NTX}$ of the proposal and $\mathsf{NTX}$ of the vote to be 5 and 3, respectively. Fig.~\ref{fig:Dynamic-Latency} and Fig.~\ref{fig:Throughput-Dynamic-Latency} depict the relationship of latency and throughput \emph{vs.} packet loss rate, respectively, for 20 different consensus protocols.

\subsubsection{ReduceCatch Advantages} Experiments demonstrate that ReduceCatch achieves lower latency and higher throughput compared to other consensus protocols relying on different communication protocols, especially when packet loss surpasses 10\%. Specifically, for the same consensus algorithm, ReduceCatch outperforms the other four communication protocols by an average latency advantage of 10.77 seconds. Furthermore, it establishes a significant lead of 44.28 seconds over the fastest protocol within Tendermint2 (utilizing CSMA-ACK). Furthermore, the ReduceCatch-based consensus protocol displays a higher average throughput of over 3.82 TPM for the same consensus algorithm under the other four communication protocols, with a 14.8 TPM advantage over the highest throughput obtained in PBFT (based on CSMA-NACK). 



\begin{figure*} 
\centering
  \subfloat[Latency]{
   \includegraphics[width=0.47\linewidth]{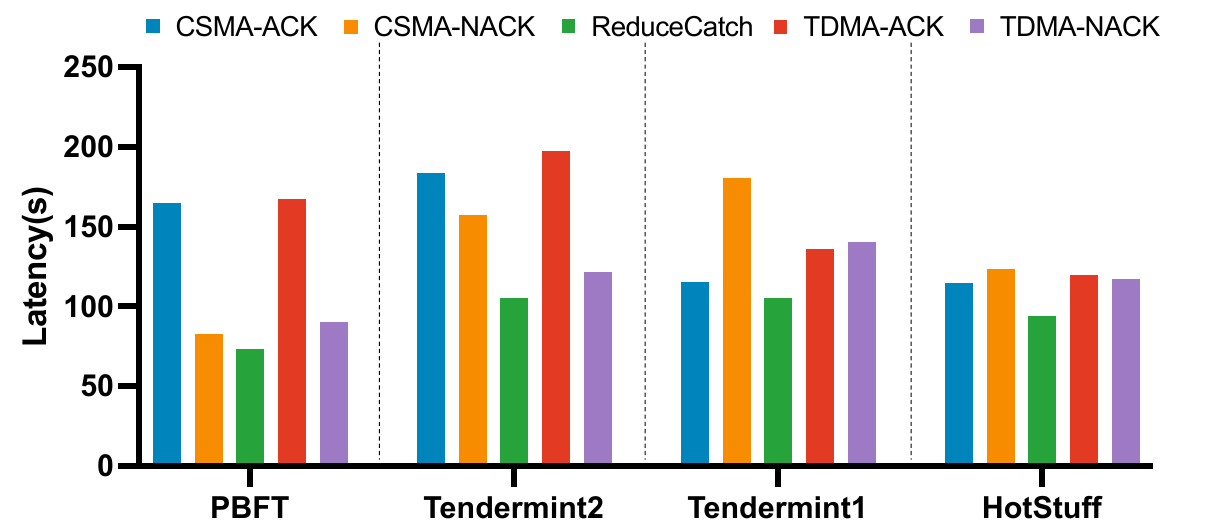}
\label{fig:Multi-Hop-Latency}}\hfill
  \subfloat[Throughput]{
    \includegraphics[width=0.47\linewidth]{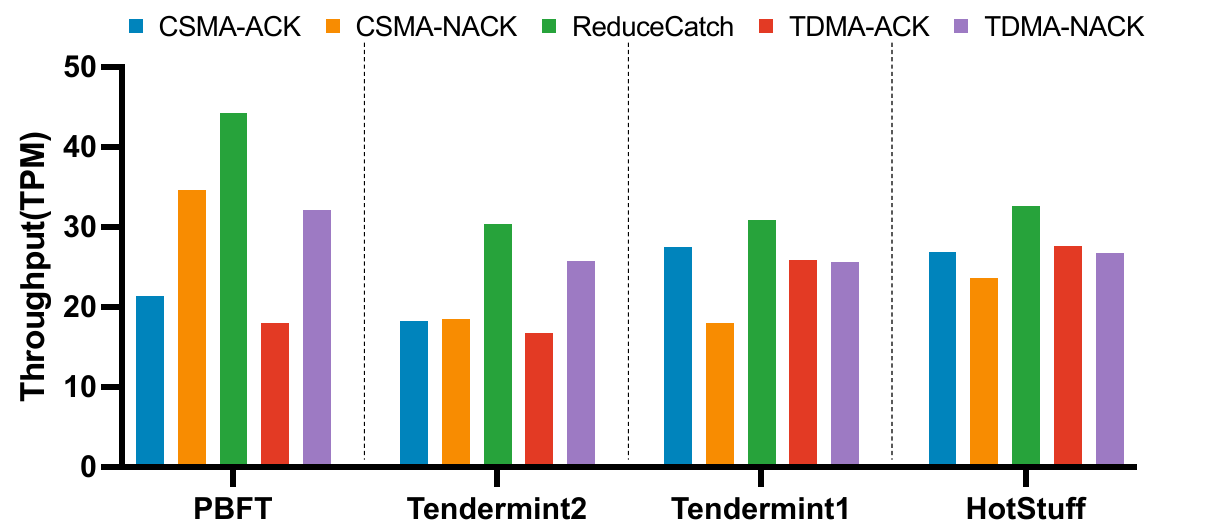}
\label{fig:Multi-Hop-Throughput}}
  \caption{The latency or throughput of four consensus algorithms under multi-hop scenarios}
  \label{fig:Multi-Hop-Evaluation} 
\end{figure*}

\subsubsection{Communication Patterns}

TDMA and ACK both have a negative impact on consensus performance, but their degree of impact varies depending on the communication pattern employed.
In consensus protocols based on N-to-N communications, ACK has a greater impact on performance, while in protocols based on 1-to-N and N-to-1 communications, TDMA dominates.
%
For PBFT (Fig.~\ref{fig:Dynamic-PBFT} and Fig.~\ref{fig:Throughput-Dynamic-PBFT}) or Tendermint1 (Fig.~\ref{fig:Dynamic-Tendermint1} and Fig.~\ref{fig:Throughput-Dynamic-Tendermint1}), the performance of TDMA-ACK (TDMA-NACK), in terms of latency and throughput, is weaker than that of CSMA-ACK (CSMA-NACK) for PBFT and Tendermint1. This aspect is also reflected in HotStuff and Tendermint2, which primarily adopt 1-to-N and N-to-1 communications, particularly when the packet loss rate increases. Notably, in Tendermint2, the performance based on CSMA-NACK is slightly weaker than that of TDMA-NACK (Fig.~\ref{fig:Dynamic-Tendermint2} and Fig.~\ref{fig:Throughput-Dynamic-Tendermint2}), primarily due to the influence of the view change phase. The reason that TDMA has a more significant impact than CSMA in dynamic networks lies in that the interaction between two nodes depends not only on network congestion but also on the time interval between them. Although TDMA has lower network congestion than CSMA, the round trip time for interaction between two nodes requires $N$ time slots, which is 10 time slots (10 seconds) in our experiment. The interaction time in CSMA depends on the timer size, which is 4.5 seconds in this experiment. As network conditions deteriorate, the impact of node interaction time on performance surpasses that of network congestion.

The 1-to-N and N-to-1 communications are better suited for adverse network environments compared to the N-to-N communications. This can be observed by comparing HotStuff and PBFT. In static networks, when the communication protocol is CSMA-ACK, HotStuff exhibits weaker performance than PBFT in terms of latency and throughput. Due to the exacerbation of channel congestion by CSMA and the increased message complexity caused by ACK, combined with the increased network dynamics, HotStuff's performance surpasses that of PBFT.


\subsubsection{View Change}
The impact of view changes in Tendermint1 versus PBFT is non-linear across all five protocols and worsens with network dynamics. Specifically, the complexity of 1-to-N or N-to-1 communications during view changes grows with network dynamism, leading to significantly higher latency in dynamic networks—a critical factor for distributed system design.
As network dynamics intensify, the view change phase in Tendermint2 exhibits weaker performance than that of HotStuff. 
As network dynamics increase, the performance of Tendermint2 becomes weaker compared to HotStuff. With CSMA-NACK, as network dynamics increase, the performance of Tendermint2 considerably weakens in comparison to HotStuff. Besides, under CSMA-ACK, Tendermint2 outperforms HotStuff, and the performance gap between the two narrows with increasing network dynamics. These two phenomena highlight the inferior stability of the view change phase in Tendermint2 compared to HotStuff. The primary reason is that Tendermint requires collecting $N$ \textit{view change} messages, which is challenging due to increasing dynamics, whereas HotStuff only needs $2f+1$ \textit{view change} messages.

\subsection{Multi-Hop}

\label{sec:evaluation:Multi-Hop}
We conduct a performance evaluation on 20 different consensus protocols in a multi-hop network. The network is divided into four clusters, each consisting of four nodes. To showcase the dynamic network performance of ReduceCatch, we set the packet loss rate to 30$\%$. Consequently, we set $\mathsf{NTX}$ of the proposal to 5 and $\mathsf{NTX}$ of the vote to 3. Our results, depicted in Figs.~\ref{fig:Multi-Hop-Latency} and~\ref{fig:Multi-Hop-Throughput}, respectively demonstrate the latency and throughput of the twenty consensus protocols in multi-hop networks.

\subsubsection{ReduceCatch Advantages}
Presented in Figs.~\ref{fig:Multi-Hop-Latency} and~\ref{fig:Multi-Hop-Throughput}, the ReduceCatch-based consensus protocol outperforms the other four protocols for the same consensus algorithm in terms of latency and throughput. Specifically, the ReduceCatch-based consensus protocol exhibits a latency improvement of more than 9.8 seconds compared to the fastest consensus protocol under the other four communication protocols. Furthermore, in HotStuff, the ReduceCatch-based consensus protocol is 20.7 seconds faster than the fastest consensus protocol (based on CSMA-ACK). In terms of throughput, the ReduceCatch-based consensus protocol is more than 3.4 TPM higher than the highest throughput among the other four communication protocols for the same consensus algorithm. In particular, in PBFT, the throughput of the ReduceCatch-based consensus protocol is 9.7 TPM higher than the highest throughput consensus protocol (based on CSMA-NACK). ReduceCatch-based consensus protocol has strong stability in multi-hop networks, due to the reduction of participating nodes in the catch phase through the reduce phase, which reduces the network scale and improves protocol performance.


\subsubsection{Communication Patterns}
In a multi-hop network, the stability of 1-to-N and N-to-1 communications is greater than that of a N-to-N communication. As shown in Figs.~\ref{fig:Multi-Hop-Latency} and~\ref{fig:Multi-Hop-Throughput}, the performance difference among the five communication protocols of PBFT or Tendermint1 is noticeable, unlike that of HotStuff and Tendermint2. Additionally, the performance comparison of PBFT and HotStuff in multi-hop networks is similar to that in dynamic networks (Section~\ref{sec:evaluation:Dynamic}), where HotStuff outperforms PBFT under CSMA-ACK and TDMA-ACK. 
This further illustrates that complex network topologies benefit from 1-to-N and N-to-1 communications over N-to-N.

\subsubsection{View Change}
Similar to the findings in Section~\ref{sec:evaluation:Dynamic}, in a multi-hop network, Tendermint1 exhibits weaker performance in terms of latency and throughput compared to PBFT. Specifically, With CSMA-NACK, Tendermint1 is 74.3 seconds slower and 16.2 TPM lower in throughput. Comparing performance under CSMA-NACK from a static single-hop network (4 nodes) to a multi-hop network (16 nodes, 30\% packet loss), the latency difference grows 6.2 times rather than doubling.
This is due to the challenges of multi-hop networks and dynamic conditions, which adversely affect the view change phase in Tendermint1.


\section{Conclusion and Future Directions}
\label{sec:conclusion}

In this paper, we present a communication protocol, named ReduceCatch (Reduce and Catch), to optimize the partially synchronous Byzantine fault-tolerant (BFT) consensus at the network layer in ad hoc wireless networks. We also establish a wireless consensus testbed to evaluate the partially synchronous BFT consensus in ad hoc wireless networks. 
Looking ahead, we plan to explore three key aspects: implementing synchronous and asynchronous BFT consensus on our testbed, examining wireless applications like satellite scenarios, and integrating embodied AI to enhance cooperation efficiency and system performance.


\newpage
\bibliographystyle{IEEEtran}
\bibliography{ref}

\end{document}